\documentclass[final,3p,times,twocolumn,sort&compress]{elsarticle}

\usepackage[caption=false]{subfig}
\usepackage{graphicx} 
\usepackage{amsmath}
\usepackage{amssymb}
\usepackage{amsthm}
\usepackage{bm}
\usepackage{graphicx} 
\usepackage{siunitx} 
\usepackage{float}
\usepackage{tikz}
\usetikzlibrary{arrows.meta}
\usepackage{cases}
\usepackage{nicefrac}
\usepackage{booktabs}
\usetikzlibrary{shadows,arrows,decorations.pathmorphing,backgrounds,positioning,fit,3d,er,calc,math}
\pgfdeclarelayer{background}
\pgfdeclarelayer{foreground}
\pgfsetlayers{background,main,foreground}

\usepackage[pagebackref]{hyperref}
\hypersetup{
    colorlinks,%
    citecolor=blue,%
    filecolor=black,%
    linkcolor=magenta,%
    urlcolor=black,%
    pdfauthor=author
}

\journal{Scripta Materialia (Accepted for Publication)}

\begin{document}
\begin{frontmatter}

\title{Temperature-activated dislocation avalanches signaling brittle-to-ductile transition in BCC micropillars}

\author[SHU]{Yang Li\fnref{equal}}%
\author[UM]{Inam Lalani\fnref{equal}}%
\author[UM]{Matthew Maron}%
\author[UM]{William Hixson}%
\author[SYSU]{Biao Wang}%
\author[UCLA]{Nasr Ghoniem\corref{cor1}}%
\ead{ghoniem@ucla.edu}
\author[UM]{Giacomo Po\corref{cor1}}%
\ead{gpo@miami.edu}
\cortext[cor1]{Corresponding authors.}

\fntext[equal]{These authors contributed equally to this work.}

\address[SHU]{School of Mechanics \& Engineering Science, Shanghai University, Shanghai 200072, China}%
\address[UM]{Department of Mechanical \& Aerospace Engineering, University of Miami, Coral Gables, FL 33146, USA}%
\address[SYSU]{School of Physics, Sun Yat-sen University, Guangzhou 510275, China}%
\address[UCLA]{Department of Mechanical \& Aerospace Engineering, University of California Los Angeles, Los Angeles, CA 90095, USA}

\date{\today}

\begin{abstract}
We carry out strain-controlled in situ compression experiments of micron-sized tungsten micropillars (W) in the temperature range 300-900~K, together with simulations of three-dimensional discrete dislocation dynamics (DDD) on the same scale. Two distinct regimes are observed. At low temperatures, plastic deformation appears smooth,  both temporally and spatially. Stress fluctuations are consistent with a Wiener stochastic process resulting from uncorrelated dislocation activity within the pillars. However, high-temperature stress fluctuations are highly correlated and exhibit features of self-organized criticality (SOC), with deformation located within well-defined slip bands. The high-temperature stress relaxation statistics are consistent with a thermally activated nucleation process from the surface. The nature of the transition between the two regimes is a manifestation of the brittle to ductile transition in BCC metals.
\end{abstract}


\end{frontmatter}

\begin{figure*}[t!]
\centering
    \subfloat[]{
    \includegraphics[trim = 1cm 0cm 1.5cm 1cm,width=0.53\textwidth]{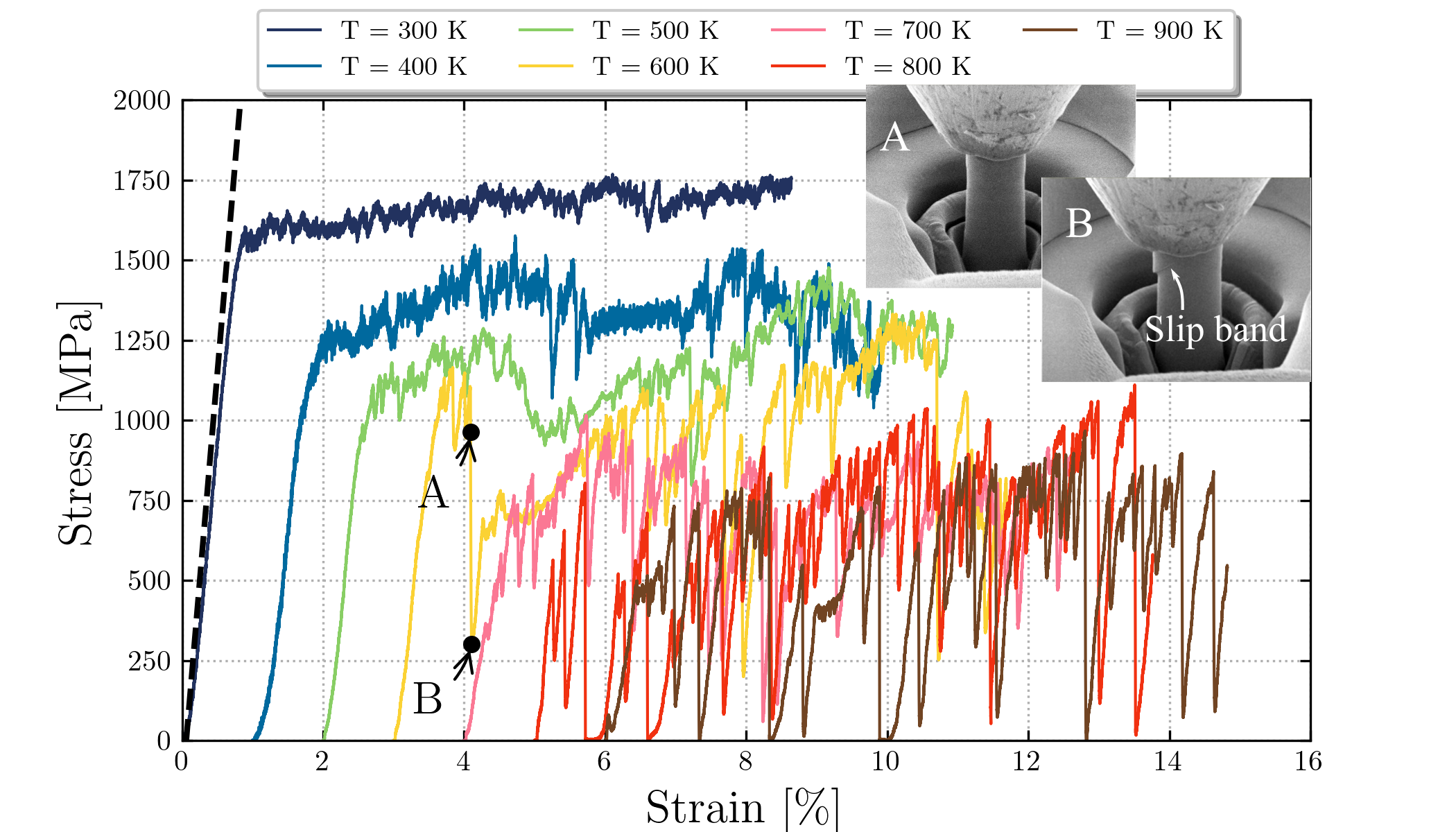}
             \label{fig:exp_ssa}} 
             \hfill
        \subfloat[]{
    \includegraphics[width=0.43\textwidth]{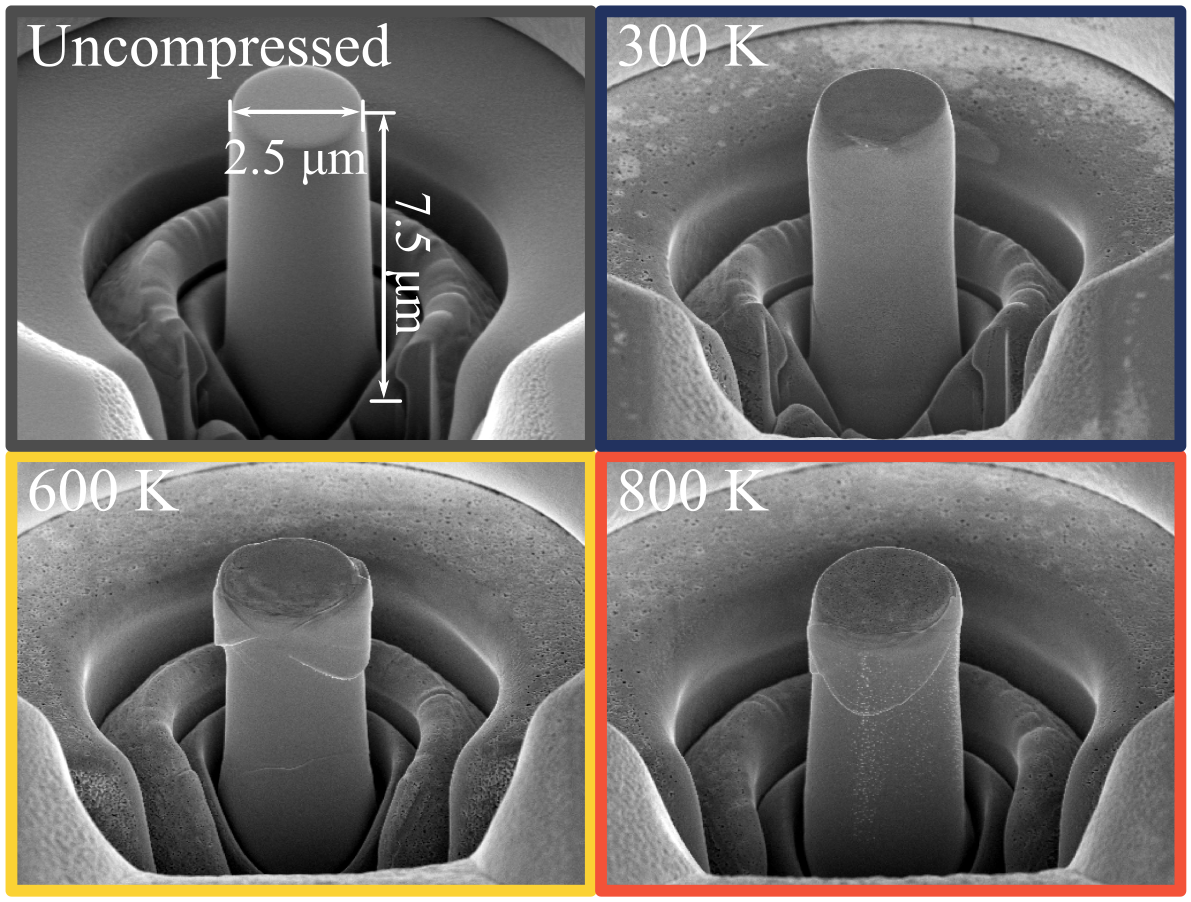}
             \label{fig:exp_ssb}} 
    \caption{ (a) Micropillar compression stress-strain curves from 300~K to 900~K.  An incremental $1\%$ strain shift is applied to each temperature for clarity. The SEM images in the insert indicate the occurrence of a large strain burst. (b) Size of representative pre-compressed pillar and post-compression SEM images of micropillars at 300~K, 600~K and 800~K.}
    \label{fig:exp_ss}
\end{figure*}

During the last two decades, acoustic emission and micropillar compression experiments have revealed that plastic deformation in crystalline materials manifests at the microscale in the form of intermittent dislocation activity \citep{weiss2001complexity,dimiduk2006scale,ispanovity2022dislocation,papanikolaou2017avalanches,weiss2003three,brinckmann2008fundamental,zaiser2008strain}. This observation suggests the intriguing hypothesis that, at a fundamental level, plasticity could be governed by the principle of self-organized criticality (SOC) \cite{bak1987self}, a hypothesis already advanced within the context of other physical phenomena, such as solar flares \cite{crosby1993frequency}, earthquakes \cite{sornette1989self}, snow avalanches, and sandpile slides \cite{miguel2001intermittent,maass2018micro}. Compared to these fields, where experimental parameters are often impossible to change,  microscale plasticity provided a unique platform to test one of the fundamental assumptions of SOC, that the cutoff of the size distribution depends only on the size of the system and not on external tunable parameters \cite{markovic2014power}. In fact, micropillar experiments appeared to defy this assumption, revealing that external parameters such as stress \cite{friedman2012statistics} and strain rate \cite{maass2018micro} affect the size distribution cutoff, thus giving rise to the alternative concept of \emph{tuned} criticality. It was proposed that the SOC-like behavior of certain crystals, such as ice, could be the limit of a continuum spectrum, with a ``mild-to-wild" transition controlled by a single ratio $R=L/\ell$ between the size of the system $L$ and an intrinsic length scale $\ell=\mu b/\tau $ determined by the overall resistance $\tau$ to dislocation motion \cite{weiss2019ice}. For example, the introduction of precipitates of high pinning strength in Al pillars was shown to inhibit strain bursts \cite{zhang2017taming}. More recently, micropillar compression tests carried out in body-centered cubic (BCC) crystals showed that the distribution of avalanche sizes either follows a truncated power law similar to face-centered cubic (FCC) crystals, but with a temperature-dependent cutoff \cite{alcala2020statistics,zhang2023coupled}, or exhibits an exponential to power law transition as temperature increases \cite{rizzardi2022mild}. 
\textcolor{black}{Specifically, it remains unclear whether temperature acts primarily as a tuning parameter that modulates the size cutoff of a universal power-law distribution, or if it drives a fundamental transition in the stochastic nature of dislocation activity.}
We posit that the role of temperature in relation to self-critical behavior in BCC metals remains not fully understood.

In this work, we show that low-temperature avalanches in W micropillars are consistent with a Wiener stochastic process resulting from uncorrelated dislocation activity within the pillars. In contrast, high-temperature stress fluctuations are highly correlated, exhibit characteristics of SOC, and are consistent with a different internal mechanism. We show that the transition between these two regimes is a universal feature of BCC micropillars and that the nature of this transition is the same as in the brittle-to-ductile transition (BDT) of BCC metals.

Micropillars were fabricated in a 99.999\%-pure $\langle 100 \rangle$-oriented ($\pm 2^\circ$) floating-zone-grown single-crystal tungsten sample, polished to a roughness of less than $10$nm.  Micropillars of nominal 2.5~\si{\mu m}  diameter and $3:1$ aspect ratio were milled using a focused ion beam (FIB), achieving a taper angle of 2.0$\pm$0.5 degrees. The micropillars were uniaxially compressed \textit{in-situ} using a Bruker PI-88 nanoindenter within a Zeiss UltraPlus SEM. The nanoindenter used a 5~\si{\mu m} flat WC probe and was equipped with an extended-range piezoelectric load cell, as well as probe and sample heating stages. The target displacement rate of the proportional-integral (PI) control loop was set to 2~nm/s, corresponding to a strain rate of $\sim 2.67\times 10^{-4}$~\si{s^{-1}}.
\textcolor{black}{The chosen strain rate is selected to balance experimental drift correction with the requirement for quasi-static driving. While low-temperature BCC plasticity is inherently slow due to screw dislocation lattice friction, the consistency between our experimental results, DDD simulations, and the weak strain rate sensitivity observed in similar BCC systems \cite{srivastava_influence_2021} suggests that the observed stochastic regimes are physical.}
\textcolor{black}{It is noted, however, that while our rate is quasi-static for W, lower rates may be required to capture the  adiabatic limit of the fluctuation distribution \cite{sparks2018nontrivial}.}
All micropillars were compressed to a total engineering strain of $\sim 10\%$ before unloading. The tests were carried out in the temperature range of 300~K to 900~K in increments of 100~K, with measurements repeated on 2-4 pillars at each temperature. The probe and sample temperatures were controlled by independent PI feedback loops, which were carefully tuned to limit the temperature fluctuation upon contact to less than 0.1~K.

\begin{figure*}[t!]
\centering
    \subfloat[]{
            \includegraphics[trim = 1cm 0cm 0.9cm 0.5cm, width=0.48\textwidth, keepaspectratio]{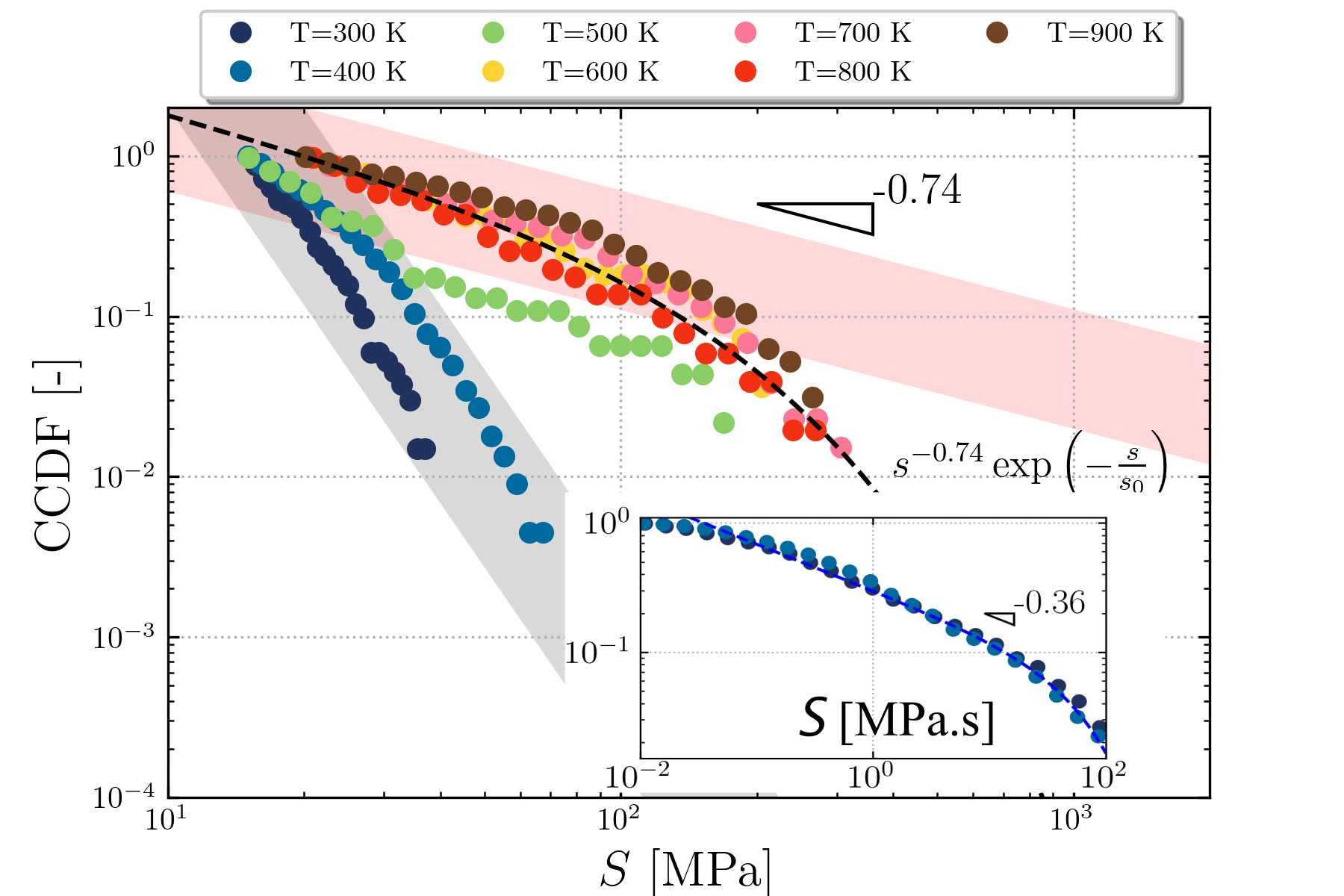}
             \label{fig:size}} 
             \hfill
    \subfloat[]{
            \includegraphics[trim = 1cm 0cm 0.9cm 0.5cm, width=0.48\textwidth, keepaspectratio]{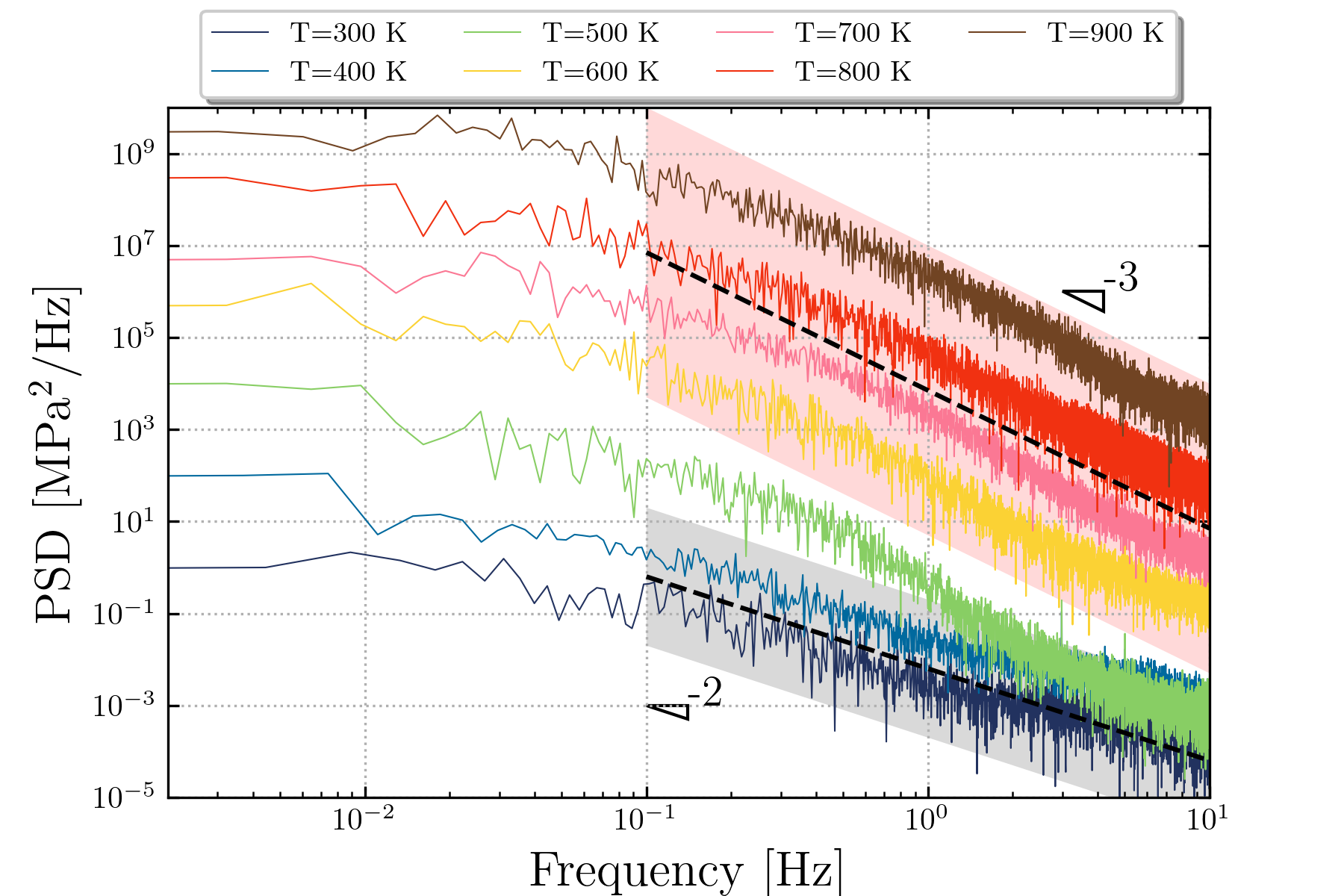}
             \label{fig:psd}}
    \caption{(a) Complementary cumulative distribution function (CCDF) of avalanche size at different temperatures. The inset shows that the \textcolor{black}{truncated} power law distribution of the ``size" $\mathcal{S}$ is consistent with a stochastic Wiener process at low temperature. (b) PSD of stress fluctuations at different temperatures. For clarity, a shift along the $y$ -axis is applied.  
    }
    \label{fig:exp}
\end{figure*}

Representative stress-strain curves of micropillar compression tests are presented in Fig.~\ref{fig:exp_ss}, together with selected pre- and post-compression SEM images. The room temperature Young's modulus $E\approx 260$~GPa and the strength $\sigma_{y}\approx 1.56$~GPa are consistent with previous measurements on W pillars of similar size \cite{schneider2009correlation,abad2016temperature}.
The overall decrease in strength with increasing temperature is a signature of the thermally activated mobility of screw dislocations in BCC crystals. The striking feature of these stress-strain curves is the transition from a smooth plastic flow to abrupt stress drops with increasing temperature. 
\textcolor{black}{This distinct transition in dislocation activity with increasing temperature can be also observed from SEM images. At 300~K, deformation is spatially homogeneous, lacking well-defined slip bands. Above 600~K, prominent planar slip traces emerge. Crystallographic slip-trace analysis indicates these bands correspond to $\{110\}$ planes, with the measured tilt-corrected inclination of $\sim 45^\circ$ matching the theoretical prediction for the $\{110\}$ families. 
}

This transition can be characterized by the distribution of the size of the dislocation avalanche. To construct avalanche statistics, we recall that the primary indicator of the dynamics of collective displacement is the plastic strain rate $\dot\varepsilon^p$, which is proportional to the stress rate $\dot{\sigma}\approx -E \dot\varepsilon^p$ when avalanches occur under a sufficiently slow displacement control. Following \cite{laurson20061,papanikolaou2012quasi}, Wiener filtering was applied to the nanoindenter load cell signal, which, upon appropriate conversion to an engineering stress rate $\dot{\sigma}$, was filtered to extract avalanche events of duration $T$ and size $S = \int_0^T \dot{\sigma} \text{d}t$. \textcolor{black}{At temperatures of 800~K and above, the displacement signal was also used to limit the duration $T$ whenever stress drops were amplified by the nanoindenter feedback loop, as detailed in the supplementary materials.} Fig.~\ref{fig:size} shows the complementary cumulative distribution functions (CCDF) of event sizes for different temperatures. A truncated power law scaling is found to capture data above 600~K, with an exponent of \textcolor{black}{$\eta \sim 0.74\pm0.04$} for the burst size. This exponent is comparable to the value observed in many pure metals (between 0.3 and 1.1), including Ni, Au, Al, Cu, Mo, and Nb \cite{dimiduk2006scale,brinckmann2008fundamental,greer2011plasticity,zhang2017taming,papanikolaou2017avalanches}. With decreasing temperature, the size of events decreases drastically, and distributions spanning less than a decade are hardly attributed to critical behavior \cite{stumpf2012critical}. A similar trend in the temperature dependence of the avalanche size distribution was also observed in other BCC micropillars, such as Nb and Mo \cite{rizzardi2022mild,zhang2023coupled}.

It should be recognized that power law \emph{size} distributions are generated by many processes, including branching processes and random walks, and therefore represent a necessary but not sufficient signature of critical phenomena \cite{watkins201625,jensen2021critical}. Due to the lack of dominant characteristic length and time scales, temporal (and spatial) \emph{correlations} are the true hallmark of critical phenomena \cite{kuntz2000noise,laurson2005power}. Compared to the avalanche size analysis, the power spectral density (PSD) of stress fluctuations offers therefore more direct evidence of correlated dislocation activity within the micropillars. The PSD of the raw stress signal is shown in Fig.~\ref{fig:psd}, while the effects of the nanoindenter feedback loop on the PSD are discussed  in the supplementary materials.  The PSD is well described by a power law $P_\sigma(f) \sim f^{-\gamma}$ at all temperatures, which implies $P_{\dot{\varepsilon}^p}(f) \sim f^{-\gamma+2}$ for the plastic strain rate. At temperatures above 600~K, a power law exponent $\gamma\sim3$ is observed, corresponding to a $\sim 1/f$ \lq\lq crackling\rq\rq  noise in the plastic deformation rate. This indicates that high-temperature plasticity within the W micropillars is dominated by highly correlated dislocation activity. Consistent with the characteristics of SOC \cite{bak1987self}, the correlated activity manifests itself in heavy-tailed avalanche size distributions, as observed in Fig.~\ref{fig:size} at high temperatures. By contrast, below 500~K the stress PSD scales as a power law with exponent $\gamma\sim2$, which is typical of a stochastic Wiener process (or a continuous Brownian motion). Note that the size CCDF of a Wiener process is a power law with exponent $1/3$ \cite{villegas2019time}, which is in agreement with the distribution of ``sizes" $\mathcal{S} = \int_0^T \sigma \text{d}t$ shown in the inset of Fig.~\ref{fig:size} at low temperatures. A further validation, that the mean squared displacement of the stress signal exhibits a linear relations with respect to time, is shown in the supplementary materials. A Wiener process represents additive white noise, that is, $P_{\dot{\varepsilon}^p}(f) \sim f^0$. Hence, low-temperature plasticity in the micropillars appears to be mediated by uncorrelated dislocation activity. This uncorrelated dislocation activity, which is an indication of negligible dislocation-dislocation interactions, also manifests itself in the lack of significant strain hardening.

\begin{figure*}[t!]
   \subfloat[]{
        \includegraphics[trim = 0.6cm 0cm 0cm 0.cm, width=\columnwidth, keepaspectratio]{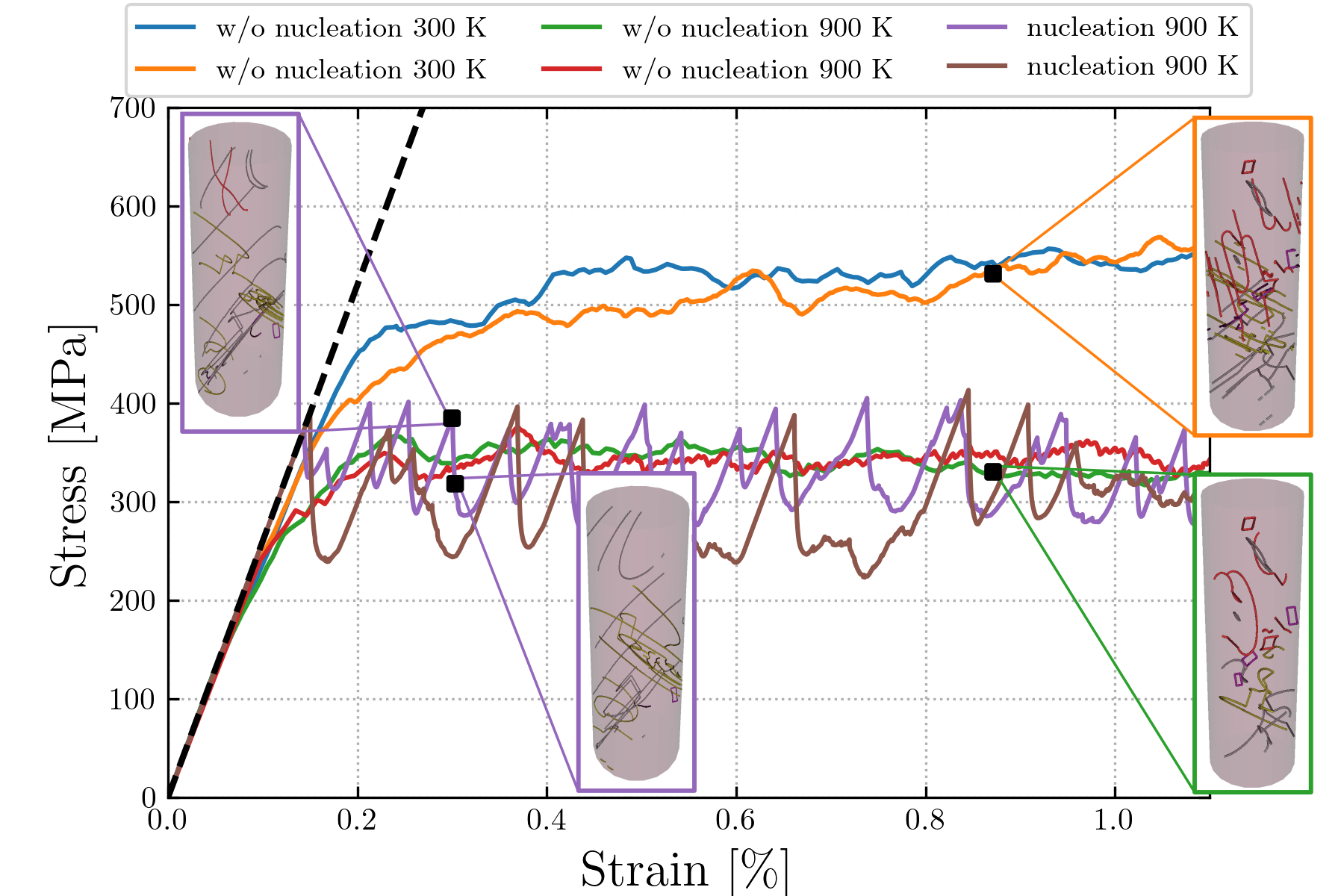}
         \label{fig:ss_sim}}  \hfill
   \subfloat[]{
        \includegraphics[trim = 0.6cm 0cm 0cm 0.cm, width=\columnwidth, keepaspectratio]{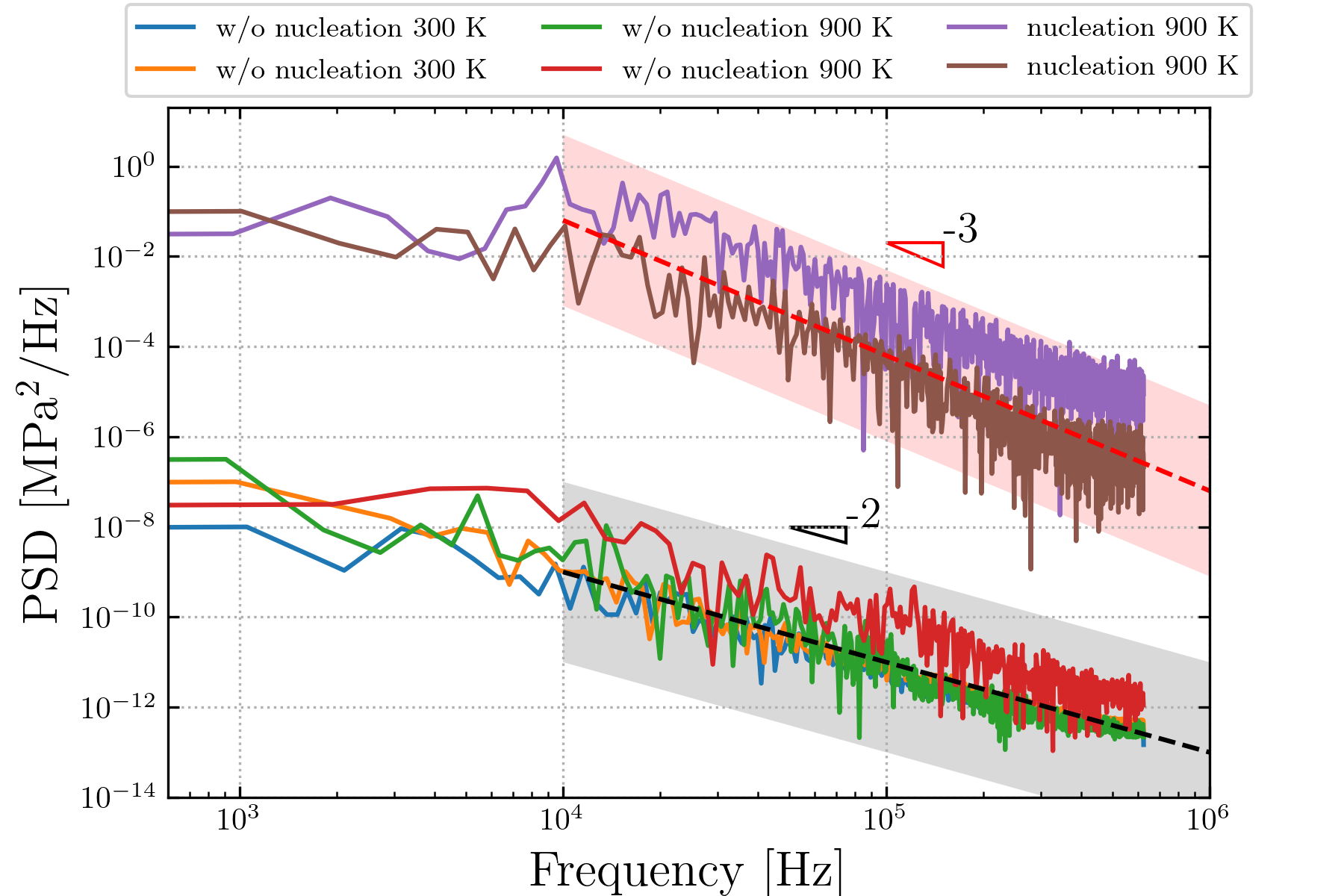}
         \label{fig:psd_sim}}
    \caption{ (a) Stress-strain curves of DDD micropillar compression simulations at 300~K and 900~K  with and without surface nucleation. Each case is simulated multiple times with different initial configurations. (b) Corresponding power spectrum of the flow stress from the simulations.
    }
    \label{fig:sim}
\end{figure*}

To elucidate the underlying mechanisms of low- and high-temperature plasticity in W micropillars, we carry out three-dimensional DDD simulations in cylindrical domains with the same dimensions and crystallographic orientations as the experimental samples. We employ the MoDELib computer package, which features a dislocation mobility law tuned to capture the thermally activated motion of screw dislocations in W \cite{po2016phenomenological,cui2018size}, a pencil-glide cross-slip model, and elastic corrections for free and fixed surface boundary effects. We assume that internal dislocation sources are the residue of the single-crystal growth process. A series of simulations are performed at 300~K and 900~K, with randomly seeded initial dislocation sources including Frank-Read sources, single-arm sources, and mixed sources. At 300~K, screw dislocation mobility is limited and many sources are activated to accommodate the imposed strain rate. Plastic flow is maintained by slow and steady propagation of long straight screw dislocations, with periodic emission of fast edge components from the source activation process, consistent with previous simulations \cite{cui2016temperature,alcala2020statistics}, as well as TEM observations \cite{pozuelo2021separation}. Representative stress-strain curves and PSD of stress fluctuations are shown in Fig.~\ref{fig:sim}. Although the flow stress depends slightly on the specific initial microstructure, at 300~K the PSD of all simulations \textcolor{black}{with only internal sources} scales as $~1/f^2$. The PSD power law exponent agrees well with our experiments, confirming that low-temperature plasticity is a superposition of the uncorrelated movement of short-lived edge and long-lived screw dislocations. As expected, our DDD simulations capture the temperature dependence of the flow stress and the emergence of a mixed and curved dislocation microstructure with increasing temperature \cite{po2016phenomenological}. 
\textcolor{black}{However, using only internal sources, the PSD $P_\sigma(f)$ still scales as $1/f^2$ at 900~K (the case without the surface nucleation), and no large stress drops are observed in the simulated stress-strain curves, in disagreement with the experimental data. We conclude that another thermally activated process is responsible for the large stress fluctuations at high temperatures. }

\begin{figure}[b!]
    \centering
    \includegraphics[width=\columnwidth, keepaspectratio]{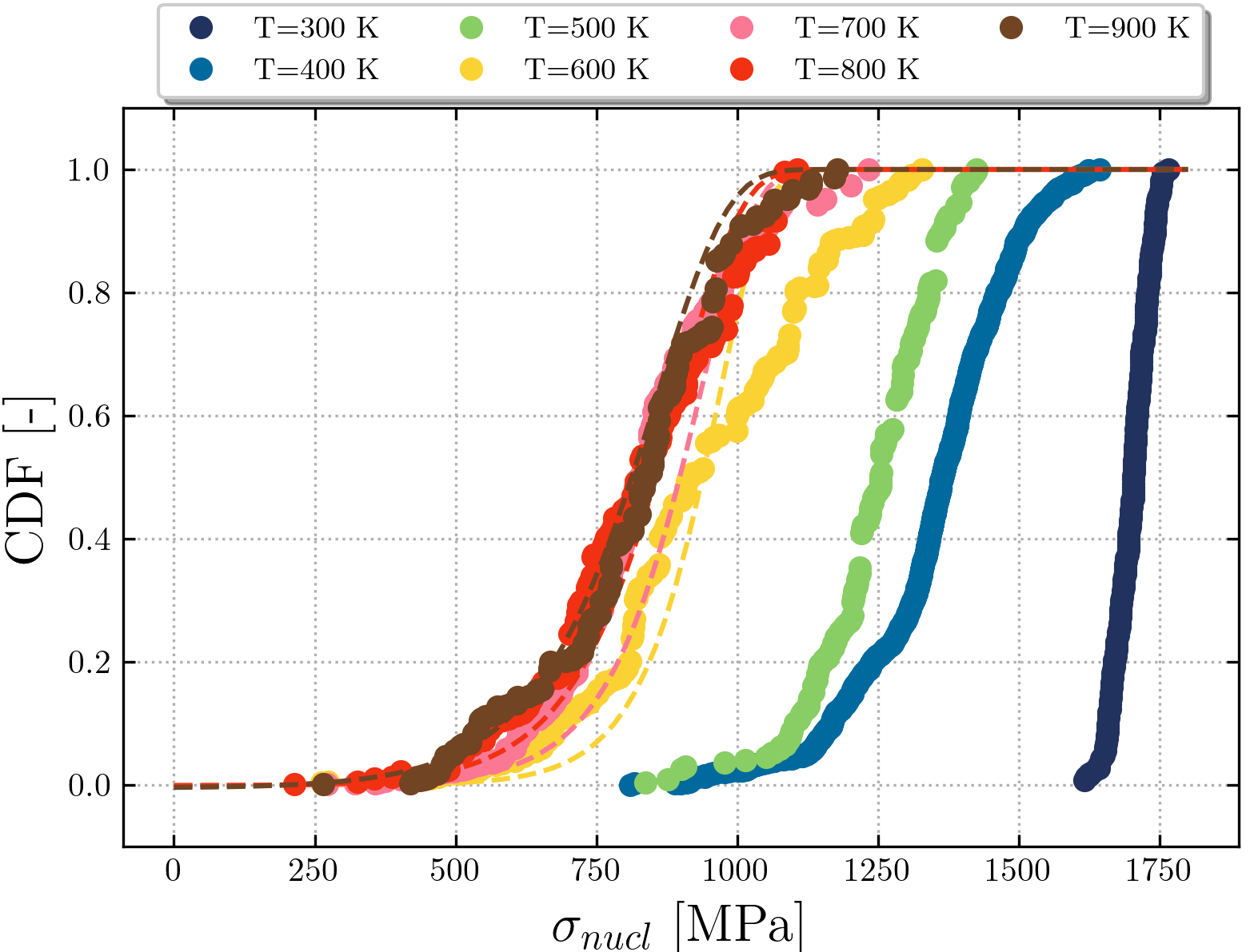}
    \caption{Cumulative probability function of the nucleation stress. The dashed lines are model predictions from the supplementary equation~11.}
    \label{fig:CDF_nucleation}
\end{figure}

To identify the nature of this process, we consider the statistics of the number of dislocations that exit a pillar during an avalanche event. The number of dislocation loops that leave a pillar within a stress drop $\Delta \sigma$ is found from the relation $\Delta \sigma = E (\Delta \varepsilon - n b/h)$, where $\Delta \varepsilon$ is the applied strain increment, $\Delta \varepsilon^p = n b/h$ is the plastic strain increment, $b$ is the magnitude of the Burgers vector along the loading direction and $h$ is the height of the pillar. Calculation details can be found in the supplementary materials. Guided by the images of deformed crystals in our experiments (see Fig. \ref{fig:exp_ss}), we assume that these dislocations $n$ are emitted near the peak stress of the event. We can then construct the CDF of the number of dislocations emitted as a function of stress and temperature, as shown in Fig.~\ref{fig:CDF_nucleation}. A model of this CDF can be developed as follows. Dislocations are emitted at a rate $\nu(\sigma,T) = N \nu_0 \exp\left( - \Delta H(\sigma) /\, k_B T\right)$, where $\nu_0$ is the attempt frequency, and $N$ is the number of source sites \cite{zhu2008temperature}.  The activation enthalpy $\Delta H(\sigma) = \Delta U_0 - m\sigma V$ depends on the shear stress $\tau=m\sigma$, where $m$ is the Schmid factor,  $\Delta U_0$ is the activation energy, and $V$ is the activation volume. Following \cite{chen2015measuring}, the cumulative distribution of dislocations emitted from all sources is $\text{CDF}(\sigma,T)=1-\text{exp}\left(-\frac{1}{E\dot\varepsilon}\int_0^\sigma \nu(\xi,T)\, \text{d}\xi\right)$, which assumes elastic loading between burst events. Using nonlinear least squares regression, we fit this function to the high-temperature experimental data in Fig.~\ref{fig:CDF_nucleation}, and find that $N\nu_0 \sim 3.6~\si{s^{-1}}$, $V \sim 8 b^3$ and $\Delta U_0 \sim 0.68$~eV.

This activation energy is lower than the value expected for the glide of screw dislocations \cite{gumbsch1998controlling}, indicating that mobility is not the only mechanism controlling the transition between uncorrelated and correlated plastic activity in the pillars. Indeed, both activation energy and the volumes are consistent with a nucleation process, as demonstrated by other experiments and molecular dynamics simulations \cite{zhu2008temperature,brochard2010elastic,chen2015measuring}. Nucleation is expected to occur from areas of stress concentration on the pillar surface, such as indenter-pillar contact points and FIB-induced nanometer-scale ledges. This was already proposed by Schneider \textit{et al.}, based on the effects of FIB re-milling in Mo micropillars of comparable size compressed at room temperature \cite{schneider2010effect}. 
\textcolor{black}{Direct experimental evidence has been further demonstrated by Maass \textit{et al.} \cite{maass2012situ,maass2018micro}. Using in-situ Laue micro-diffraction, they clearly revealed that dislocations are generated and stored at the sample-indenter interface during Ni-pillar compression prior to the breakaway stress.}
The synergistic effects of sufficient dislocation mobility and increased nucleation rate also explain the high temporal and spatial correlation features of the avalanche events observed at high temperatures. This is because, given enough mobility, nucleation can act to relax local stress concentrations by emitting dislocation arrays. These coherent dislocations, emitted from a localized source, sweep the crystal on adjacent planes and within a short time window, thus forming localized slip bands and large stress drops. The coincidence of temporal and spatial events is observed in our results, where slip bands become evident after large stress drops (e.g., inset in Fig.~\ref{fig:exp_ss}). Because proper flow stress is never achieved before a large nucleation event takes place, at temperatures above 600~K this process produces saw-tooth stress-strain profiles, as seen in Fig.~\ref{fig:exp_ss}. Individual source exhaustion caused by local stress relaxation typically leads to the activation of multiple nucleation sources and to the formation of the corresponding slip bands during a single test. This is further validated by our DDD simulations. By incorporating the nucleation model above with parameters fitted from the experiments, our DDD simulations successfully capture large stress drops, highly localized plastic deformation, and the $1/f^3$ PSD scaling (Fig.~\ref{fig:sim}), as well as the power-law size distribution of stress drops.

Our results indicate that low-temperature avalanches are the manifestation of uncorrelated plastic activity dominated by low-mobility screw dislocations, showing no indication of correlated events or critical behavior. A transition to a regime of highly correlated events is observed with increasing temperature. 
\textcolor{black}{While the observed transition temperature is not an intrinsic material constant but scales with the system size and the strain rate  \cite{zhang2023coupled}, the similar deformation characteristics of different BCC micropillars at comparable homologous temperatures \cite{schneider2009correlation,abad2016temperature} suggest that this transition is a universal feature of BCC metals.}

\textcolor{black}{
This transition arises from the combined effects of an increasing dislocation nucleation rate and enhanced dislocation mobility with increasing temperature. Taken together, these factors make it difficult not to draw a parallel with the well-known brittle-to-ductile (BDT) transition in bulk BCC metals \cite{gumbsch1998controlling}.
The similarities between the two transitions are striking. First, in both cases the transition occurs when two mechanisms act synergistically: dislocations are nucleated at a sufficiently high rate from stress concentrators—internal cracks in bulk samples and surface defects in micropillars—and their mobility becomes sufficiently high to accommodate the imposed deformation rate. Second, the characteristic transition temperatures and activation energies associated with these processes are comparable. In pure tungsten, the bulk BDT typically occurs in the range of $\sim 400$–$700$~K \cite{gumbsch2003brittle,faleschini2007fracture}, with reported activation energies of $\sim 0.2$–$0.5$~eV \cite{gumbsch1998controlling,ast2018brittle}. Our finding of $\Delta U_0 \sim 0.68$~eV for surface nucleation is consistent with this energy scale, suggesting a shared rate-limiting mechanism. Based on these observations, our work suggests that the transition from smooth, uncorrelated deformation to avalanche-controlled collective behavior observed here signals a fundamental change in dislocation dynamics in BCC metals, corresponding to the brittle-to-ductile transition in bulk samples. Although further work is needed to firmly establish this connection, linking these two transitions has important practical implications, as it would enable the screening of BDT properties in BCC alloys using avalanche statistics obtained from inexpensive, high-throughput, small-scale tests.}

\section*{Acknowledgments}
Y.L. acknowledges the support of the National Natural Science Foundation of China under Grant No. 12302279. G.P. acknowledges the support of the US DOE through grants number DE-SC0024675 and DE-SC0024401, and the US NSF through CAREER grant number 2340174 with the University of Miami. The support of the U.S. Department of Energy at UCLA is acknowledged under DE-SC0018410:0008. B.W. acknowledges the support of the NSFC Original Exploration Project under Grant No. 12150001, and the Guangdong Provincial Key Laboratory of Extreme Conditions under 2023B1212010002.

\bibliographystyle{elsarticle-num}
\bibliography{Wpillar.bib}

\end{document}